\hoffset -0.0 true cm
\hsize 16 true cm         %horizontal size
\vsize 22 true cm         %vertical size
%\rm                         %type of font
%
%%%%%%%%%%%%%%%    TEXT FORMATS    %%%%%%%%%%%%%%%
%
\def\doublespace {\smallskipamount=6pt plus2pt minus2pt
                  \medskipamount=12pt plus4pt minus4pt
                  \bigskipamount=24pt plus8pt minus8pt
                  \normalbaselineskip=24pt plus0pt minus0pt
                  \normallineskip=2pt
                  \normallineskiplimit=0pt
                  \jot=6pt
                  {\def\smallskip {\vskip\smallskipamount}}
                  {\def\medskip   {\vskip\medskipamount}}
                  {\def\bigskip   {\vskip\bigskipamount}}
                  {\setbox\strutbox=\hbox{\vrule 
                    height17.0pt depth7.0pt width 0pt}}
                  \normalbaselines}
%
%     1-1/2-spaced manuscript definition
%
\def\pprintspace {\smallskipamount=4pt plus1pt minus1pt
                  \medskipamount=9pt plus2pt minus2pt
                  \bigskipamount=16pt plus4pt minus4pt
                  \normalbaselineskip=14pt plus0pt minus0pt
                  \normallineskip=1pt
                  \normallineskiplimit=0pt
                  \jot=4pt
                  {\def\smallskip {\vskip\smallskipamount}}
                  {\def\medskip   {\vskip\medskipamount}}
                  {\def\bigskip   {\vskip\bigskipamount}}
                  {\setbox\strutbox=\hbox{\vrule 
                    height8.5pt depth3.5pt width 0pt}}
                  \parskip 4pt
                  \normalbaselines}
\doublespace 
%\pprintspace
%\magstep 1
\message {Format established}
%
%Title page
%
\vbox{
\hbox{}
\vskip1.0truein
\centerline{\bf  Radio Continuum Imaging of
High Redshift  Radio Galaxies}
%\centerline{\bf Images, Core Identifications, Rotation Measures, and
%the P-D-Z Relationships to z $>$ 4}
\vskip1.2truein
\centerline {\bf C.L. Carilli$^{1,2}$, H.J.A. R\"ottgering$^2$, 
R. van Ojik$^2$, G.K. Miley$^2$}
\vskip0.1truein
\centerline{\bf W.J.M. van Breugel$^3$}
\vskip 0.2truein
\centerline{$^1$National Radio Astronomy Observatory, PO Box O,
Socorro, NM 87801}
\centerline{~$^2$Leiden Observatory, Postbus 9513, 2300 RA, Leiden, 
The Netherlands}
\centerline {~$^3$Lawrence Livermore National Laboratories,  L-413
PO Box 808, 7000 East Ave., Livermore, CA, 94550}
\vskip1.0truein
\centerline{Submitted to: The Astrophysical Journal (Supplement Series), 
December 19, 1995}
}

\vfill\eject
\centerline {\bf ABSTRACT}  

We present sensitive
radio continuum images at high resolution of 37 radio galaxies at z $>$ 2.
The  observations were made with 
the  Very Large Array (VLA) at 4.7 GHz and 
8.2 GHz, with typical resolutions of 0.45$''$ and 0.25$''$, respectively.
Images of total and polarized intensity, and spectral index, are
presented. Values for total and polarized intensity, and values
of rotation measures, are tabulated for the hot spots in each source.
The positions of the
radio nuclei are tabulated, along with a variety of other
source parameters. 

Analysis of the polarization data reveals large rotation measures (RMs)
towards six sources. We argue that the RMs are
due to magnetized, ionized gas local to the radio sources. The 
magnitude of the RMs are in excess of 1000 rad m$^{-2}$  (rest frame)
for these sources. Drawing an analogy to a class
of lower redshift radio galaxies with extreme RMs, 
we speculate that these sources
may be at the centers of dense, X-ray emitting cluster atmospheres.

\vfill\eject

\smallskip
\centerline {\bf 1. Introduction}
\smallskip

Prior to 1989  only a single radio galaxy had been identified beyond
z~=~2.  Since then there have been over 60 galaxies identified beyond
z~=~2 ({\sl cf.} McCarthy 1993),
including eight beyond z~=~3, and one beyond z~=~4 (Lacy {\sl et al.} 1994).
Standard cosmologies  dictate that the universe at z~=~2 
is only 20$\%$ its present age\footnote{$^1$}{We use q$_o$~=~0.5, and 
h~$\equiv$ ~H$_o$/(100 km sec$^{-1}$ Mpc$^{-1}$).}.
During this epoch the space density of 
luminous active galaxies was a few hundred times larger than today 
(Dunlop and Peacock 1990). High redshift radio galaxies are 
extremely luminous,  and spatially extended, in many wavebands. Hence,
the study of high redshift radio galaxies  continues
to play an important role in our understanding of the cosmic environment 
at large redshift.

The most effective method for finding high redshift radio
galaxies is by selecting for steep spectrum radio sources with 
faint optical counterparts ({\sl cf.} Chambers, Miley, and van Breugel
1990, Miley and Chambers 1989, McCarthy {\sl etal.} 
1990, R\"ottgering 1993, van Ojik 1995). 
Although such sources are  targeted on the basis
of their radio continuum properties, subsequent study has focused
almost exclusively on optical and near-infrared observations.
Radio continuum studies of these sources have remained cursory.
Sensitive, high resolution, polarimetric radio imaging 
provides important information
on these sources and their environments in many ways, including: 
(i) identification of the location of the active
nucleus, (ii) study of the cosmological evolution of radio source structure
({\sl cf.} Barthel and Miley 1988),
(iii) tests of quasar-radio galaxy unification schemes as a function of 
redshift ({\sl cf.} Antonucci 1993), 
(iv) searching for extreme rotation measures and ultra-steep spectrum regions
(Carilli, Owen, and Harris 1994, Carilli 1995, Tribble 1993), 
(v) determination of the projected magnetic field morphology,  and 
(vi) detection of radio jets, and study of correlations between jet 
sided-ness and depolarization, or spectral index, asymmetries
({\sl cf.} Garrington {\sl et al.} 1989, 1991, Laing 1988a).
All of these parameters probe radio source physics, as well as the
physical environments of the sources.
High resolution images of a large sample of sources can also be used
to search for small scale gravitational lensing events (scales $<$ 0.5$''$)
towards the extended structures in the radio sources -- thereby constraining
the space density of small galaxies ($\le$ 0.1 L$_*$) at intermediate
redshifts (Carilli {\sl etal.} 1994, Carilli 1995, Kochanek and Lawrence 1990,
Law-Green etal. 1995).

The discovery of the radio-optical `alignment effect' for 
high redshift radio galaxies
(Chambers, Miley, and van Breugel 1987, McCarthy {\sl etal.} 
1987), establishes a clear 
relationship between the radio and optical emission from 
such galaxies.  Detailed comparative 
studies in the different spectral regimes are 
essential for understanding the various processes involved in this
phenomenon.
High resolution radio images provide a unique data-set for comparison 
with high resolution optical images
in order to study the radio-optical alignment effect on
sub-kpc scales ({\sl cf.} Miley {\sl et al.} 1992, Carilli, Owen,
and Harris 1994, Best, Longair, and R\"ottgering 1996, Chambers {\sl
et al.} 1996). Such small scale alignments provide important constraints
on shocks, scattering, and/or jet-induced star 
formation as possible causes for this  
enigmatic phenomenon (McCarthy 1993). 

This paper presents multifrequency images of the total and
polarized radio continuum emission from 37 radio galaxies at z $>$ 2.
Section 2 describes the sample, and section 3 describes the
observations and data reduction. The images and basic source
parameters are presented in Section 4. In section 5 we present 
preliminary physical analysis of a few basic parameters, such
as core properties,  and the origin of large rotation measures. 

\smallskip
\centerline{\bf 2. The Sample}
\smallskip

The Leiden observatory has, for some years, been performing a 
systematic search for radio galaxies at high redshift under the
direction of G. Miley
(R\"ottgering  etal. 1996, R\"ottgering {\sl et al.} 1995, van
Ojik 1995, R\"ottgering {\sl et al.} 1994,  R\"ottgering 1993, 
Chambers etal 1990, Chambers{\sl et al.}
1987, Chambers 1988, Chambers {\sl et al.} 1996).
One of the primary goals of this search is to increase the number of
known sources 
at high redshift in order to study the redshift evolution of radio source
properties.  We have selected a sub-sample of sources
from the Leiden sample of high redshift radio galaxies, supplemented with
sources  taken from the literature, for follow-up high 
resolution polarimetric imaging with the Very Large Array.
The sample was originally defined as an optical magnitude
limited sample for an imaging program with 
the Hubble Space Telescope, with the criteria of: z $\ge$ 2
and optical magnitude~ R $\le$ 23.5.  

The sources are listed in Table 1. The source name 
and redshift are listed in Columns 1 and 2. The catalog
in which the radio source was first identified is listed in
column 3. References to these catalogs are given
in the notes to the table. Columns 4, 5, 6, 7, and 8 list the source 
R magnitude, the maximum extent of the radio
source,  and integrated flux densities at 1.5 GHz, 4.7 GHz, and 8.2 GHz, 
respectively. Column 9 lists the reference for the optical identification and
redshift determination.

The nomenclature `radio galaxy' for these sources (as opposed to `radio
quasar') is based on the observation of
relatively narrow emission lines from these sources, typically
$\approx$ 1000 km sec$^{-1}$, and in many cases on the fact that
the optical emission is spatially  extended, although this latter
conclusion is unclear for some of the optically 
fainter sources in the sample.
It should be emphasized that to date no direct evidence exists indicating
that the optical
continuum emission from these sources is stellar in origin.

\smallskip
\centerline{\bf 3. Observations and Data Reduction}
\smallskip

Observations were made using the Very Large Array 
(Napier, Thompson, and Ekers 1983) in its
A (27 km) configuration on March 18 and 19, 1994. 
We employed two frequencies in the 5 GHz band of the VLA (4535 MHz and
4885 MHz), and two frequencies in the 8 GHz band (8085 MHz and 8335 MHz).
Bandwidths at all frequencies were 50 MHz. 
Each source was observed for 20 min at 8 GHz and 10 min at 5 GHz.
The theoretical noise is 25 $\mu$Jy at 8 GHz, and 40 $\mu$Jy at 5 GHz.
In most cases this noise level is achieved on the final image.

All data  were processed using the Astronomical Image Processing
System (AIPS). Standard gain calibration 
was done using 3C 286, and checked with 
scans of 3C 48. We estimate the uncertainty in 
absolute fluxes to be $\le$ 2$\%$.
An important limitation for large angular scale sources is the
lack of the shortest spacings in the data. The  A array short spacing
limit implies a maximum size on which we have information 
of about 10$''$ at 5 GHz and 6$''$ at 8 GHz. This limitation will have a 
negligible effect when considering spectra of bright, small components, such as
the hot spots and cores of the radio sources, but will have a large
effect when considering  extended structures in big sources.

The on-axis antenna  polarization response terms were determined
using multiple scans of  the  calibrator 0746+483 over a large range in 
parallactic angle. Absolute linear polarization position angles were 
measured  using two scans of 3C 286 separated in time by 6 hours.  
From the difference in solutions between the two scans on 3C 286 
we estimate an uncertainty in the observed position angles due to 
calibration (ie. in addition to those dictated by
signal-to-noise) of about 2$^o$ at all frequencies.
This minimum error sets a  minimum RM magnitude which can be 
measured between 5 GHz and 8 GHz
of about 25 rad m$^{-2}$. Below we adopt a 4$\sigma$ limit to
RM magnitude of 100 rad m$^{-2}$ as a reliable RM detection.

The calibrated data were then edited and self-calibrated using 
standard procedures (Perley 1988) to improve image dynamic range.
For the faintest sources the first self-calibration iteration at 8 GHz
involved phase self-calibration using a model derived from the 5 GHz
data. Natural weighting of the 
gridded visibilities was employed in the final
imaging stage in order to maximize sensitivity. Images were deconvolved
using the CLEAN algorithm as implemented in the AIPS task MX. 
The FWHM of the Gaussian restoring beams are listed in Table 2.
Images of the three Stokes polarization parameters, I, Q, and U
were synthesized, and all images were CLEANed down to the level of 
2.5 times the theoretical RMS on the image. 

For total intensity analysis the data from the two 
frequencies per band were combined, and
we adopt the mean frequency in each band (8210 MHz and 
4710 MHz). Spectral index images were generated 
by convolving the 8 GHz image with the Gaussian restoring beam of the
5 GHz image, then checking for small astrometry differences possibly 
introduced in the phase boot-strap procedure, or during
self-calibration, by looking at the positions
of the hot spots and core. 
Spectral index, $\alpha$, 
is defined as a function of surface brightness, I$_\nu$,
at  frequency, $\nu$, as: I$_\nu$ $\propto$ $\nu^{\alpha}$.
Astrometric differences between images at the
two frequencies were typically $<$ 0.1$''$, and always $<$ 0.2$''$.
Spectral index values were calculated only for regions with surface
brightnesses $>$ 4$\sigma$ at each frequency, where $\sigma$ is the
measured off-source RMS on an image.

Rotation measures were derived using  position angles 
for the polarized intensity
from three frequencies: 4535 MHz, 4885 MHZ, and 8200 MHz.
Rotation measures were derived for the polarized emission from the hot
spots in each lobe, as listed in Table 3, by fitting a quadratic
function in wavelength to the observed position angles at the
three frequencies. Fitting was done only for hot spots with 
polarized intensities $>$ 4$\sigma$. The maximum rotation measure
that can be measured is set by assuming less than $\pi$/2 rad rotation between
the two frequencies in the 5 GHz band, implying a
limit of:~ RM $<$ 5200 rad m$^{-2}$.
The RM values quoted in Table 3 were derived
using the observed frequencies. If the Faraday `screens'
are local to the source, then a factor of (1+z)$^2$ is required in the 
calculation of RM. We return to this point in Section 5.

\smallskip
\centerline{\bf 4. Images and Observed Parameters}
\smallskip

Images of total intensity at 4.7 GHz and 8.2 GHz are shown in 
Figure 1, along with images of spectral index 
between 4.7 GHz and 8.2 GHz. Also shown
are images of polarized intensity at 5 GHz, along
with position angles for the electric field vectors of the
polarized emission.

The sources have been observed in either J2000 or B1950 coordinates, as
dictated by the original identification work. We have chosen to
preserve the position equinox from the original identification work
in our current observations 
in order to facilitate comparison between these images and
existing data at other bands.  The position equinox is noted on 
each image.  

One important question that can be addressed with these data 
are the positions of
the active nuclei. We identify the nuclei as unresolved components
located between the outer-most components, and having spectral
indices which are as flat as, or flatter than,  any other source component.
In Table 2 we list the position of the nucleus in each  source in which a
nucleus was detected. Column 1 lists the source name, and column
two lists the nuclear position. Again, we have preserved positional 
equinox from the original optical identification work for ease of
comparison with previous work at optical and other wavebands. 
The equinox of the  position is indicated
in each case. Columns 3 and 4 list the 
flux densities of the nuclei at 8.2 GHz and  the spectral index 
between 4.7 and 8.2 GHz, respectively.
Column 5 lists the core fraction, as defined by the 
ratio of the core flux density to the total flux density at a rest frame
frequency of 20 GHz, as derived by extrapolating the integrated source spectrum
and the core spectrum to a rest frame frequency of 20 GHz for each
source. In many cases the nuclear identification is clear. In some cases
there is no nucleus detected, or the detection is unclear. In particular,
there are a number of sources which show central unresolved components
which have steep spectra  (spectral indices $\le$ -1), 
and/or two compact central components with comparable spectral indices. 
In these cases a `U' is placed before the core flux in Table 2
to denote uncertainty in identification.

Source parameters pertaining principally to  the hot spots
in each source are listed in Table 3. 
The table is organized such that parameters for the southern-most
hot spot in each source are in columns 2 - 5, while those for the
northern-most hot spot are in columns 6 - 10. Columns
2 - 5 give the southern hot spot peak surface brightness at 4.7 GHz, the
spectral index between 4.7 GHz and 8.2 GHz, the hot spot fractional 
polarization at 4.7 GHz and 8.2 GHz (matched resolutions), and the 
observed rotation measure at the hot spot position, respectively. 
Columns 6 - 10 give the corresponding numbers for the northern hot spots.
%Column 12 lists the ratio `R',
%of total flux of the southern regions of the source to 
%that of the  northern regions. Column 13 lists the ratio `Q',
%of the length  of the southern arm (= distance from the nucleus), to 
%that of the northern arm.

\smallskip
\centerline{\bf 5. Analysis}
\centerline{\bf 5.1 Morphological Classifications, Hot Spots, and Cores}
\smallskip

We should emphasize that the observations presented herein are at
a high frequency when redshifted into the rest frame of the 
sources (15 GHz to 30 GHz), and at high spatial resolution. 
Hence these data are well suited for studying the higher surface brightness,
flatter spectrum regions of the sources, such as hot spots and cores, but
less so  for studying the extended structures in the
sources, {\sl ie.} the radio lobes.

Most of the sources in the sample have typical edge-brightened
`Fanaroff-Riley Class II' morphologies (Fanaroff and Riley 1974), 
either having obvious hot spots at the extremities of the
radio source, or appearing as simple doubles. About 57$\%$ of the sources
show multiple hot spots on at least one side. Of the multiple hot spot sources,
about half the sources show double hot spots
(0140-257, 0406-244, 0508+606, 0748+134, 1232+397, 
1545-234, 1744+183, 1901+480, 2025-218, 2105+236, 2139-292, 2202+128), 
while in the others the hot
spot regions are  more complex (0015-229, 0156-252,
0448+091, 0828+193, 1113-178, 1138-262, 1410-001, 1436+157, 1809+407).
The parameters listed in Table 3 pertain to the brightest hot spot
in each lobe for each source.
Multiple hot spots are a common phenomenon in FRII sources
({\sl cf.} Laing 1988b, Black {\sl et al} 1992). A
likely physical interpretation of multiple radio hot spots is
that the jet has changed direction by a small amount on
a timescale $<<$ the source lifetime,
either due to a change in the ejection axis from the central engine, or
from a jet-cloud interaction along the course of the jet (Lonsdale and Barthel
1986, Carilli, Perley, and Dreher 1988, Cox {\sl etal.} 1992, van Ojik
etal 1996).

Making the standard minimum energy assumptions (Miley 1980) we derive
typical minimum pressures in the hot spots in these high redshift
sources ranging from 1.0$\times$10$^{-9}$ dyne cm$^{-2}$ to 
10.0$\times$10$^{-9}$ dyne cm$^{-2}$, 
with corresponding magnetic fields ranging from 200 $\mu$G
to 600 $\mu$G. For comparison,
the magnetic fields in the hot spots of the closest
ultra-luminous radio galaxy Cygnus A (3C 405)
are about 200 $\mu$G, as determined using both the minimum energy
assumption, and through observation of synchrotron self-Compton
X-ray emission from the hot spots (Harris, Carilli, and Perley 1994).

Five of the sources are small (14$\%$ of the full sample),
and can be considered compact steep spectrum (CSS) sources 
with sizes $\le$  10 h$^{-1}$ kpc
({\sl cf.} Fanti {\sl et al.} 1990, O'dea {\sl etal}.
1992). Of the five CSS sources,
three appear as small FRII's  (1345+245, 1324-262, and 0744+464), 
one is unresolved (0030-219),  and the fifth is a core
dominated source with very bent, stubby `arms' (2251-089). 
For comparison, the sample of steep spectrum radio loud quasars
at z $\ge$ 1.5 observed by Lonsdale, Barthel, and Miley  (1993)
showed a larger fraction (27$\%$) of CSS sources.

The most unusual looking sources in the sample are 0015-229
and 1138-262. These sources
appear as a string of knots, or `Beads-on-a-String'.  
In both cases the spectral index distribution 
does not conform to the standard FRII character
of having the flattest spectrum regions (besides the nucleus itself) 
associated with the hot spots 
at the lobe extremities. The spectra of the knots in these sources steepen
with distance from the center. These sources might be  `frustrated
jets' propagating through a dense, clumpy medium. Such a model has
been hypothesized for some CSS sources (see Fanti {\sl etal.} 1990).
The important difference for 1138-262 and 0015-229 is that the dense
medium must exist on 
a scale of 60 h$^{-1}$ kpc, about an order of magnitude larger than
for CSS sources. In the case of 1138-262 a
possible indication of a dense environment is 
the extreme rotation measure values  observed. This source
has the largest RM  in the sample (-6250 rad m$^{-2}$ in the
rest frame). Detailed investigation of 1138-262, 
including deep optical broad and narrow band imaging and 
spectroscopy, is in progress (Pentericci etal in preparation).

Most of the sources have radio core identifications (Table 2). The median
core-fraction for these sources at a rest frame frequency 
of 20 GHz is 2$\%$. For comparison, the two most luminous
radio galaxies below  z $\le$ 0.5 are
Cygnus A (3C 405) and 3C 295. 
These two sources have total radio 
luminosities similar to the high redshift sources published herein.
For these two sources the  core fractions at a rest frame
frequency of 20 GHz are: 1.5$\%$ and 0.5$\%$, respectively 
(Carilli {\sl etal.} 1991, Perley and Taylor 1991).

About  one third  of the nuclei listed in Table 2 
have a spectral index $\le$ -1.  Lonsdale, Barthel, and 
Miley (1993) also found that many of the high redshift
quasars in their study have steep spectrum cores, 
although the fraction of sources with steep spectrum cores in
their sample is not discussed. 
Lonsdale {\sl etal.} (1993) and Ramana  {\sl etal.} (1996) present a model 
for steep spectrum cores in high redshift radio sources
which is based on the substantial radio K-corrections involved:
observations at
5 GHz and  8 Ghz are sampling the core spectra at rest frame frequencies of
20 GHz to 30 GHz. They propose that the dominant component in the
core-jet  may not
be synchrotron self-absorbed at these high frequencies, thereby setting a 
lower limit of about 1 mas to the typical size of the dominant core 
component. This model can be tested  by multifrequency VLBI imaging of these
sources.

\smallskip
\centerline{\bf 5.2 Rotation Measures}
\smallskip

An important parameter from this study is that of the rotation measure
of the polarized emission from the radio source. Extensive observation
of lower redshift radio galaxies has shown a  correlation between 
large rotation measures and cluster environment: all sources 
located at the centers of dense, X-ray emitting cluster atmospheres show
large amounts of Faraday rotation (Taylor, Barton, and Ge 1994). 
Indeed, Taylor {\sl etal.} find a clear correlation
between cluster core thermal electron 
density derived from X-ray observations, and 
magnitude of Faraday rotation derived from radio observations. 
An important point is that this correlation is  independent
of radio source luminosity and morphological class, 
and hence is a  probe of cluster
properties, and not radio source properties. The implication is that
the hot cluster gas must be substantially magnetized, with field strengths of
order a few $\mu$G. 

An important question to address for the sources in the current study 
is whether the observed rotation measures are Galactic 
in origin, or are caused by a magneto-ionic medium 
local to the source itself ({\sl cf.} Pedelty {\sl etal.} 1989).
Besides the obviously different physical conclusions 
that would be reached in the two cases, there is the additional 
factor of (1+z)$^2$ which must be included in determining the RM magnitude
in the case of a screen local to the source.
We believe that for most sources
with detected rotation measures the Faraday `screen' local to 
the source, for the following reasons.
First and foremost, for most sources with detected rotation measures we find 
significant changes in the  RM values for hot spots
separated  by typically $\le$ 10$''$.
Changes larger than  100 rad m$^{-2}$ on scales less than a few arcseconds
are observed. Such large gradients on small scales
are difficult to model via a Galactic screen (Leahy 
1987). Second, the data presented herein are only sensitive to fairly large
observed-frame rotation measures (100 rad m$^{-2}$), which in themselves
are atypical of all but very low latitude Galactic lines-of-sight
(Simard-Normandin, Kronberg, and Button  1981).
And third, there are a number of sources which show significant 
depolarization between 8.2 GHz and 4.7 GHz 
at matched resolutions, indicative of 
local gradients in Faraday rotation across the sources.

Are high rotation measures common in high redshift radio sources?
To be more secure in the detection of large intrinsic RMs, we adopt the
strict standards of: polarized intensity $\ge$ 5$\sigma$ and
fractional polarization $\ge$ 1.5$\%$ at all frequencies, and
an observed RM lower limit of 100 rad m$^{-2}$.
Of the 37 sources in the sample, 6 meet these criteria:
0406-244, 1138-262, 1324-262, 1436+157, 1809+407, 2105+236, and
2202+128. Hence,  the implied rest-frame rotation measures are 
$\ge$ 1000 rad m$^{-2}$ for 19$\%$ of the sources. We consider this a lower
limit to the fraction of large RM sources  in the sample for the
following reasons. First, for the sources with no polarized emission, it
is possible that the lack of polarized emission is due to 
depolarization by gradients in a Faraday screen, 
and hence that RM values are in fact 
very large. And second, for sources with polarized flux and
low observed RM values  it could be that
the high surface brightness regions studied herein
do not sample the large RM regions. 

Overall, we can only set a lower limit of about 19$\%$ to the
fraction of sources with intrinsic rotation measures 
$\ge$ 1000 rad m$^{-2}$ in the sample. Drawing the analogy to 
lower redshift sources with extreme RMs, the simplest interpretation
is that  the  large RMs in the high redshift sources  result from 
a dense, `cooling-flow' cluster-type environment for the sources. 
Of course, we cannot rule-out the possibility that the large
RMs are produced by different physical phenomena for the high redshift
sources than the low redshift sources. An important observations would 
be to obtain deep X-ray images of the extreme RM radio galaxies 
at high redshift to search for the (hypothetical) 
cluster thermal emission. Such observations
are at the limit of current instrumentation ({\sl cf.} Crawford and 
Fabian 1993, Worrall {\sl etal.} 1994). Such observations would be
particularly significant in the light of recent debate over the
redshift evolution of the luminosity function for X-ray clusters ({\sl cf.} 
Edge {\sl etal.}  1990,  Gioia {\sl etal.} 1990, 
Luppino and Gioia 1995, Ebeling {\sl etal.} 1995), and could also
test the hypothesis that cooling flow cluster atmospheres play a
fundamental role in the formation of giant elliptical galaxies at
high redshift (Nulsen and Fabian 1995). 

A interesting alternative is to hypothesize that in some cases the large
RM screens are neither Galactic nor associated with the source, but 
cosmologically intervening material (Wolfe, Lanzetta, and
Oren 1992, Oren and Wolfe 1995).  A possible test of this idea would
be to search for faint, gas rich objects in the vicinity of the 
radio galaxy that have substantially lower redshifts than the radio galaxy.

\vskip 0.3truein

We would like to thank P. McCarthy, A. Ramana, B. McNamara, 
M. Bremer, and F. Owen for useful discussions. 
We acknowledge support by a programme subsidy provided by the Dutch
Organization for Scientific Research (NWO). CLC acknowledges support
from a NOVA research fellowship, and from the AXAF science center at
the Smithsonian Astrophysical Observatory under NASA contract NAS8-39073.
The work by WvB was performed at IGPP/LLNL under 
the auspices of the US Dept. of Energy under contract
No. W-7405-ENG-48. The National Radio Astronomy Observatory is a facility
of the National Science Foundation, operated under cooperative
agreement by Associated Universities, Inc..
This research made use of the NASA/IPAC Extragalactic Data Base (NED)
which is operated by teh Jet propulsion Lab, Caltech, under contract
with NASA.

\vfill
\eject

\parindent 0pt

\centerline{\bf References}
%

%Alexander, P. and Leahy J.P. 1987, {\sl M.N.R.A.S.}, 225, 1.

Antonucci, R.A. 1993, {\sl A.R.A.A.}, 31, 473.

Baldwin, J.E. Boysen, R., Hales, S.E.G., Jennings, J.E., Waggett,
P.C., Warner, P., Wilson, D. 1985, {\sl M.N.R.A.S.}, 217, 717.

Barthel, P.D. and Miley, G.K. 1988, {\sl Nature}, 333, 319.

%Barthel, P.D. 1989, {\sl Ap.J.} 336, 606.

Becker, R.H., White, R.L., and Edwards, A.L. 1991, {\sl Ap.J. (Supp)}, 75, 1.

Best, P.N., Longair, M.S., and R\"ottgering, H.J.A. 1996, {\sl M.N.R.A.S.
(letters)}, 280, L9.

%Begelman, M.C. 1996, in {\sl  Cygnus A: Study of a Radio Galaxy},
%eds. Carilli, C.L. and Harris, D.E., Cambridge Univ. Press, Cambridge, p. 209.

Black, A.R., Baum, S.A., Leahy, J.P., Perley, R.A., Riley, J.M., and 
Scheuer, P.A. 1992, {\sl M.N.R.A.S.}, 256, 186.

Bolton, J.G., Savage, A., and Wright, A.E. 1979, {\sl
Aust. J. Phys. Astroph. Supp.}, 46, 1.

%van Breugel, W. and McCarthy, P.  1989, in {\sl ESO Workshop on Extranuclear
%Activity in Galaxies}, eds. E.J. Meurs and R.A. Fosbury, p. 227.

Carilli, C.L., Perley, R.A., Dreher, J.W., and Leahy, J.P. 1991, {\sl Ap.J.},
383, 554.

Carilli, C.L. and Barthel, P.D. 1996, {\sl A\&A Reviews}, 7, 1.

Carilli, C.L. 1995, {\sl A\&A}, 298, 77.

Carilli, C.L., Owen, F., and Harris, D. 1994, {\sl A.J.}, 107, 480.

Carilli, C.L., Perley, R.A., and Dreher, J.W. 1988, {\sl Ap.J. (letters)},
334, L73.

Chambers, K.C., Miley, G.K., and van Breugel, W.J.M. 1987, {\sl
Nature}, 329, 604.

Chambers, K.C. 1988, {\sl Ph.D. Thesis, Johns Hopkins University}.

Chambers, K.C., Miley, G.K., and van Breugel W.J.M. 1990, {\sl Ap.J.},
363, 21.

Chambers, K.C., Miley, G.K., van Breugel, W.J.M., and Huang, J.-S.
1996, {\sl Ap.J.}, in press.

Cox, C.L., Gull, S.F., and Scheuer, P.A. 1991, {\sl M.N.R.A.S.}, 
252, 558.

Crawford, C.S. and Fabian, A.C. 1993, {\sl M.N.R.A.S.}, 260, 15p

%Daly R. 1994, Ap.J., 426, 38.

%Dreher, J.W., Carilli, C.L., and Perley, R.A. 1987,\ \ {\sl Ap.J.}, {\bf 316},
%611. 

Douglas, J.N., Bash, F., Torrence, G.W., Wolfe, C. 1980, {\sl The Univ. of
Texas Publications in Astro.}, 17, 1.

Dunlop, J.S. and Peacock, J.A. 1990, {\sl M.N.R.A.S.}, 247, 19.

%Eales, S.A., Rawlings, Steve, Puxely, Phil, Rocca-Volmerange, B., 
%and Kuntz, K. 1993, {\sl Nature}, 363, 140.

Eales, S.A., Rawlings, Steve, Dickinson, Mark, Spinrad, H., Hill, G.J.,
and Lacy, M. 1993, {\sl Ap.J.}, 409, 578

Eales, S.A. and Rawlings, Steve 1993 {\sl Ap.J.}, 411, 67

Ebeling, H., B\"ohringer, H., Briel, U., Voges, W., Edge,
A., Fabian, A., Allen, S. and Huchra, J.  1995, preprint.

Edge, A.C., Stewart, G.C., Fabian, A.C., and Arnaud, K.A. 1990, 
{\sl M.N.R.A.S.}, 245, 559.

%Eilek, J. and Shore, S. 1989, {\sl Ap.J.},  342, 187.

Fanaroff, B.L. and Riley, J.M. 1974, {\sl M.N.R.A.S.}, {\bf 167}, 31p.

Fanti, C., Fanti, R., O'Dea, C.P., and Schilizzi, R.T. 1990, {\sl The
Dwingeloo Workshop on Compact Steep Spectrum and GHz Peaked Spectrum
Radio Sources}, Istituto di Radioastronomy, Bologna, Italy.

Garrington, S.T., Conway, R.G., and Leahy, J.P. 1991, {\sl M.N.R.A.S.},
250, 171.

Garrington, S.T., Leahy, J.P., Conway, R.G., and Laing, R.A. 1988, 
{\sl Nature}, 331, 147.

Gioia, I.M., Henry, J.P., Maccacaro, T., Morris, S.L., Stocke, J.T., 
and Wolter, A. 1990, {\sl Ap.J. (letters)}, 356, L35.

Gregory, P. and Condon, J. 1991, {\sl Ap.J. (Supp)}, 75, 1011.

%Gopal-Krishna and Wiita, P.J. 1991, {\sl Ap.J.} 373, 325.

%Herbig, T. and Readhead, A. 1992, {\sl Ap.J. (Supp.)}, 81, 83.

%Kapahi, V.K. 1989, {\sl A.J.}, 97, 1.

%Killeen, N.E., Bicknell, G.V., and Ekers, R.D. 1986, {\sl Ap.J.},
%302, 306.

Hales, S.E.G., Waldram, E.M., Rees, N., and Warner, B.J. 1995,
{\sl M.N.R.A.S.}, 274, 447.

Harris, D.E., Carilli, C.L., and Perley, R.A. 1995, {\sl Nature}, 367, 713.

Kochanek, C.S. and Lawrence, C.R. 1990, {\sl A.J.}, 99, 1700.

%Krolik, J.H. and Chen, W. 1991, {\sl A.J.}, 102, 1659.

Lacy, M., Miley, G.K., Rawlings, Steve, Saunders, R., Dickinson, M.,
Garrington, S., Maddox, S., Pooley, G., Steidel, C., Bremer, M.N.,
Cotter, G., van Ojik, R., R\"ottgering, H.J.A., and Warner, P.
1994, {\sl M.N.R.A.S.}, 271, 504.

Laing, R.A. 1988a, {\sl Nature}, 31, 149.

Laing, R.A. 1988b, in {\sl Hot Spots in Extragalactic Radio Sources},
eds. K. Meisenheimer and H.-J. Roser, Springer-Verlag, Heidelberg,
p. 27.

%Laing, R.A., Riley, J.M., and Longair, M.S. 1983, {\sl M.N.R.A.S.}, 204, 151.

Law-Green, J.D., Eales, S.A., Leahy, J.P., Rawlings, S., and Lacy, M.
1995, {\sl M.N.R.A.S.}, 277, 995.

Large, M.I., Mills, B.Y., Little, A.G., Crawford, D., and Sutton, J.M. 1981,
{\sl M.N.R.A.S.}, 194, 693 

Leahy, J.P. 1987, {\sl M.N.R.A.S.}, 226, 433.

%Legg, T.H. 1970, {\sl Nature}, 226, 65.

Lonsdale, C.J., Barthel, P.D., and Miley, G.K. 1993, {\sl Ap.J. (supp)},
87, 63.

Lonsdale, C.J. and Barthel, P.D. 1986, {\sl A.J.} 92, 12.

Luppino, G.A. and Gioia, I.M. 1995, {\sl Ap.J. (letters)}, 445, L77

McCarthy, P.J. 1993, {\sl A.R.A.A.}, 31, 639

McCarthy, P.J., Kapahi, V.K., van Breugel, W.J.M., Persson, S.E., 
Ramana Athrea,  and Subrahmanya, C.R. 1996,
{\sl Ap.J.  (supp)}, in press.

McCarthy, P.J., Kapahi, V.K., van Breugel, W.J.M. and Subrahmanya, C.R. 1990,
{\sl A.J.}, 100, 1014.

McCarthy, P.J., van Breugel, W.J.M., Kapahi, V.K., and Subrahmanya, C.R.
1991, {\sl A.J.}, 102, 522.

%McCarthy, P.J., van Breugel, W.J.M., and Kapahi, V.K.
%1991b, {\sl Ap.J.}, 371, 478.

McCarthy, P.J. 1991, {\sl A.J.}, 102, 518.

McCarthy, P.J., van Breugel, W.J.M., Spinrad, H., and Djorgovski,
S. 1987, {\sl Ap.J. (letters)}, 321, L29.

Miley, G.K. 1980, {\sl A.R.A.A.}, 18, 165.

Miley, G.K. and Chambers, K.C. 1989, in {\sl ESO Workshop on Extranuclear
Activity in Galaxies}, eds. E.J. Meurs and R.A. Fosbury, p. 43.

Miley, G.K., Chambers, K.C., van Breugel, W.J., and Macchetto, F.
1992, {\sl Ap.J. (letters)}, 401, L69.

Napier, P.J., Thompson, A.R., and Ekers, R.D. 1983, {\sl Proc.
I.E.E.E.},  71, 1295.

%Neff, S.G., Hutchings, J.B., and Gower, A.C. 1989, {\sl A.J.}, 97, 1291.

%Nesser, M.J., Eales, S.A., Law-Green, J.D., Leahy, J.P., Rawlings, S.
%1995, {\sl Ap.J.} 451, 76.

%Nilsson, K., Valtonen, M.J., Kotilainen, J., and Jaakkola, T. 1993,
%{\sl Ap.J.}, 413, 453.

Nulsen, P.E.J. and Fabian, A.C. 1995, {\sl M.N.R.A.S.}, 277, 561.

O'Dea, C.P., Baum, S.A., Stanghellini, C., Dey, A., van Breugel, W., Duestua,
S. and Smith, E. 1992, {\sl A.J.}, 104, 1320.

%Oort, M.J., Katgert, P., Steerman, F.W., and Windhorst, R.A. 1987, 
%{\sl A\&A}, 179, 41.

van Ojik, R., R\"ottgering, H.A., Carilli, C.L.,  Miley, G.K., 
Bremer, M., and Macchetto, F. 1995, {\sl A\&A} in press.
 
van Ojik, R. 1995, {\sl Ph.D. Thesis}, University of Leiden.

Oren, A.L. and Wolfe, A.M. 1995 {\sl Ap.J.}, 445, 624.

Pedelty, J.A., Rudnick, L., McCarthy, P.J., and Spinrad, H. 1989,
{\sl A.J.}, 97, 647.

Perley, R. and Taylor, G. 1991, {\sl A.J.}, 101, 1623.

Perley, R.A. 1988, in {\sl  Synthesis Imaging Workshop}, 
eds. R.A. Perley, F. Schwab, and A. Bridle, NRAO, Virginia, p. 287.

Ramana Athrea, Kapahi, V.K., and McCarthy, P. 1996, in {\sl Extragalctic 
Radio Sources: Proceedings of  IAU Symposium No. 175}, eds. C. Fanti
and R. Ekers, Kluwer, Dordrecht.

Rees, N. 1990, {\sl M.N.R.A.S.}, 244, 233

R\"ottgering, H.J.A. 1993, {Ph.D. Thesis,  Leiden University}.

R\"ottgering, H.J.A., Lacy, M., Miley, G.K., Chambers, K.C., and Saunders, R. 
1994, {\sl A\&A (Supp)}, 108, 79.

R\"ottgering, H.J.A., van Ojik, R.J., Miley, G.K., Chambers, K., van 
Breugel, W., and de Koff, S. 1996, {\sl A\&A}, in press.

R\"ottgering, H.A, Miley, G.K., Chambers, K.C., and Macchetto, F. 
1995, {\sl A\&A (Supp)}, 114, 51.

Simard-Normandin, M., Kronberg, P.P., and Button, S. 1981, {\sl Ap.J. (Supp.)},
45, 97.

%Singal, A.K. 1993, {\sl M.N.R.A.S.}, 263, 139.

%Subrahmanian, K. and Swarup, G. 1990, {\sl M.N.R.A.S.}, 247, 237.

Taylor, G.B., Barton, E.J., and Ge, J.-P. 1994, {\sl A.J.}, 107, 1942.

%Taylor, G. 1991, {\sl Ph.D. Thesis, UCLA}.

Tribble, P. 1993, {\sl M.N.R.A.S.}, 261, 57.

%Wardle, J. and Miley, G.K. 1974, {\sl A\&A}, 30, 305

%Wellman, G. and Daly, R. 1995, in {\sl  Cygnus A: Study of a Radio Galaxy},
%eds. Carilli, C.L. and Harris, D.E., Cambridge Univ. Press,Cambridge, p. 215.

Vigotti, M., Grueff, G., Perley, R.A., Clark, B.G., and Bridle, A.H.
1989, {\sl A.J.}, 98, 419.

Wolfe, A.M., Lanzetta, K.M., and Oren, A.L. 1992, {\sl Ap.J.}, 388, 17

Worrall, D.M., Lawrence, C.R., Pearson, T.J., and Readhead, A.C. 1994,
{\sl Ap.J.}, 420, L17.

\vfill\eject

\parindent 10pt
\parskip 12.0pt
\centerline {\bf FIGURE CAPTIONS}

Figures 1 to 39: Images of total and polarized intensity
for the sample of high redshift radio galaxies
in Table 1. For each source there is one page of  four images:
total intensity at 4.7 GHz (upper left)  and 8.2 GHz (upper right)
at full resolution, plus 
spectral index between these two frequencies at the resolution
of the 4.7 GHz image (lower right) and polarized intensity at 4.7
GHz  (lower left).
The intensity contour levels are a geometric progression in 2$^{1/2}$, which
implies a factor 2 change in surface brightness every two contours
(negative contours included). The surface brightness of the
first contour level is indicated in the caption to each image, as are values 
for the FWHM of the Gaussian restoring beams. In all cases the
major axis of the restoring beam is oriented north-south.
The peak surface brightness in each image is also given in the caption.
The polarized intensity images also show line-segments 
indicating the observed position angles for
the electric field vectors. The contour levels for the spectral index
images are: -3.0, -2.8, -2.6, -2.4, -2.2, -2.0, -1.8, -1.6, -1.4,
-1.2, -1.0, -0.8, -0.6, -0.4, -0.2, and 0. 
The spectral index greyscale ranges from -3 to 0.

Figure 1: Radio images of 0015-229. At 4710 MHz the FWHM of the
restoring beam is 0.85$''$ $\times$ 0.44$''$, while at 8210 MHz the
beam is 0.47$''$ $\times$0.26$''$. At 4710 MHz the 
first contour level in the total
intensity image is 0.125 mJy beam$^{-1}$ and the peak surface
brightness is 17.9 mJy beam$^{-1}$. The corresponding numbers 
for the 4710 MHz polarized intensity image
are 0.12 mJy beam$^{-1}$ and 1.07 mJy beam$^{-1}$. 
At 8210 MHz the first contour level in the total
intensity image is 0.075 mJy beam$^{-1}$ and the peak surface
brightness is 7.83 mJy beam$^{-1}$.

Figure 2: Radio images of 0030-219. At 4710 MHz the FWHM of the
restoring beam is 0.84$''$ $\times$ 0.44$''$, while at 8210 MHz the
beam is 0.50$''$ $\times$0.26$''$. At 4710 MHz the 
first contour level in the total
intensity image is 0.125 mJy beam$^{-1}$ and the peak surface
brightness is 80.7 mJy beam$^{-1}$. The corresponding numbers 
for the 4710 MHz polarized intensity image
are 0.12 mJy beam$^{-1}$ and 0.21 mJy beam$^{-1}$. 
At 8210 MHz the first contour level in the total
intensity image is 0.075 mJy beam$^{-1}$ and the peak surface
brightness is 41.7 mJy beam$^{-1}$.

Figure 3: Radio images of 0140-257. At 4710 MHz the FWHM of the
restoring beam is 0.88$''$ $\times$0.44$''$, while at 8210 MHz the
beam is 0.48$''$ $\times$0.27$''$. At 4710 MHz the 
first contour level in the total
intensity image is 0.125 mJy beam$^{-1}$ and the peak surface
brightness is 24.7 mJy beam$^{-1}$. The corresponding numbers 
for the 4710 MHz polarized intensity image
are 0.12 mJy beam$^{-1}$ and 1.36 mJy beam$^{-1}$. 
At 8210 MHz the first contour level in the total
intensity image is 0.075 mJy beam$^{-1}$ and the peak surface
brightness is 10.9 mJy beam$^{-1}$.

Figure 4: Radio images of 0156-252. At 4710 MHz the FWHM of the
restoring beam is 0.90$''$ $\times$0.44$''$, while at 8210 MHz the
beam is 0.49$''$ $\times$0.25$''$. At 4710 MHz the 
first contour level in the total
intensity image is 0.175 mJy beam$^{-1}$ and the peak surface
brightness is 59.5 mJy beam$^{-1}$. The corresponding numbers 
for the 4710 MHz polarized intensity image
are 0.12 mJy beam$^{-1}$ and 3.70 mJy beam$^{-1}$. 
At 8210 MHz the first contour level in the total
intensity image is 0.075 mJy beam$^{-1}$ and the peak surface
brightness is 25.5 mJy beam$^{-1}$.

Figure 5: Radio images of 0200+015. At 4710 MHz the FWHM of the
restoring beam is 0.54$''$ $\times$0.45$''$, while at 8210 MHz the
beam is 0.31$''$ $\times$0.25$''$. At 4710 MHz the 
first contour level in the total
intensity image is 0.125 mJy beam$^{-1}$ and the peak surface
brightness is 21.4 mJy beam$^{-1}$. The corresponding numbers 
for the 4710 MHz polarized intensity image
are 0.12 mJy beam$^{-1}$ and 4.04 mJy beam$^{-1}$. 
At 8210 MHz the first contour level in the total
intensity image is 0.125 mJy beam$^{-1}$ and the peak surface
brightness is  7.0 mJy beam$^{-1}$.

Figure 6: Radio images of 0211-122. At 4710 MHz the FWHM of the
restoring beam is 0.68$''$ $\times$0.44$''$, while at 8210 MHz the
beam is 0.39$''$ $\times$0.26$''$. At 4710 MHz the 
first contour level in the total
intensity image is 0.125 mJy beam$^{-1}$ and the peak surface
brightness is 18.8 mJy beam$^{-1}$. The corresponding numbers 
for the 4710 MHz polarized intensity image
are 0.12 mJy beam$^{-1}$ and 2.28 mJy beam$^{-1}$. 
At 8210 MHz the first contour level in the total
intensity image is 0.075 mJy beam$^{-1}$ and the peak surface
brightness is  7.7 mJy beam$^{-1}$. 

Figure 7: Radio images of 0214+183. At 4710 MHz the FWHM of the
restoring beam is 0.48$''$ $\times$0.46$''$, while at 8210 MHz the
beam is 0.28$''$ $\times$0.26$''$. At 4710 MHz the 
first contour level in the total
intensity image is 0.150 mJy beam$^{-1}$ and the peak surface
brightness is 58.4 mJy beam$^{-1}$. The corresponding numbers 
for the 4710 MHz polarized intensity image
are 0.12 mJy beam$^{-1}$ and 4.99 mJy beam$^{-1}$. 
At 8210 MHz the first contour level in the total
intensity image is 0.090 mJy beam$^{-1}$ and the peak surface
brightness is 25.7 mJy beam$^{-1}$. 

Figure 8: Radio images of 0316-257. At 4710 MHz the FWHM of the
restoring beam is 0.90$''$ $\times$0.44$''$, while at 8210 MHz the
beam is 0.48$''$ $\times$0.25$''$. At 4710 MHz the 
first contour level in the total
intensity image is 0.125 mJy beam$^{-1}$ and the peak surface
brightness is 53.7 mJy beam$^{-1}$. The corresponding numbers 
for the 4710 MHz polarized intensity image
are 0.12 mJy beam$^{-1}$ and 0.65 mJy beam$^{-1}$. 
At 8210 MHz the first contour level in the total
intensity image is 0.075 mJy beam$^{-1}$ and the peak surface
brightness is 20.9 mJy beam$^{-1}$. 

Figure 9: Radio images of 0406-244. At 4710 MHz the FWHM of the
restoring beam is 0.87$''$ $\times$0.44$''$, while at 8210 MHz the
beam is 0.51$''$ $\times$0.25$''$. At 4710 MHz the 
first contour level in the total
intensity image is 0.125 mJy beam$^{-1}$ and the peak surface
brightness is 55.2 mJy beam$^{-1}$. The corresponding numbers 
for the 4710 MHz polarized intensity image
are 0.12 mJy beam$^{-1}$ and 4.05 mJy beam$^{-1}$. 
At 8210 MHz the first contour level in the total
intensity image is 0.075 mJy beam$^{-1}$ and the peak surface
brightness is 22.7 mJy beam$^{-1}$. 

Figure 10:
Radio images of 0417-181. At 4710 MHz the FWHM of the
restoring beam is 0.72$''$ $\times$0.44$''$, while at 8210 MHz the
beam is 0.42$''$ $\times$0.25$''$. At 4710 MHz the 
first contour level in the total
intensity image is 0.125 mJy beam$^{-1}$ and the peak surface
brightness is 46.5 mJy beam$^{-1}$. The corresponding numbers 
for the 4710 MHz polarized intensity image
are 0.12 mJy beam$^{-1}$ and 0.38 mJy beam$^{-1}$. 
At 8210 MHz the first contour level in the total
intensity image is 0.075 mJy beam$^{-1}$ and the peak surface
brightness is 24.5 mJy beam$^{-1}$. 

Figure 11: 
Radio images of 0448+091. At 4710 MHz the FWHM of the
restoring beam is 0.51$''$ $\times$0.45$''$, while at 8210 MHz the
beam is 0.30$''$ $\times$0.29$''$. At 4710 MHz the 
first contour level in the total
intensity image is 0.125 mJy beam$^{-1}$ and the peak surface
brightness is 3.62 mJy beam$^{-1}$. The corresponding numbers 
for the 4710 MHz polarized intensity image
are 0.12 mJy beam$^{-1}$ and 0.29 mJy beam$^{-1}$. 
At 8210 MHz the first contour level in the total
intensity image is 0.075 mJy beam$^{-1}$ and the peak surface
brightness is 0.94 mJy beam$^{-1}$. 

Figure 12: 
Radio images of 0508+606. At 4710 MHz the FWHM of the
restoring beam is 0.48$''$ $\times$0.44$''$, while at 8210 MHz the
beam is 0.29$''$ $\times$0.26$''$. At 4710 MHz the 
first contour level in the total
intensity image is 0.125 mJy beam$^{-1}$ and the peak surface
brightness is 12.5 mJy beam$^{-1}$. The corresponding numbers 
for the 4710 MHz polarized intensity image
are 0.12 mJy beam$^{-1}$ and 0.86 mJy beam$^{-1}$. 
At 8210 MHz the first contour level in the total
intensity image is 0.075 mJy beam$^{-1}$ and the peak surface
brightness is  4.5 mJy beam$^{-1}$. 

Figure 13: Radio images of 0731+438. At 4710 MHz the FWHM of the
restoring beam is 0.46$''$ $\times$0.43$''$, while at 8210 MHz the
beam is 0.28$''$ $\times$0.25$''$. At 4710 MHz the 
first contour level in the total
intensity image is 0.150 mJy beam$^{-1}$ and the peak surface
brightness is 62.9 mJy beam$^{-1}$. The corresponding numbers 
for the 4710 MHz polarized intensity image
are 0.12 mJy beam$^{-1}$ and 3.62 mJy beam$^{-1}$. 
At 8210 MHz the first contour level in the total
intensity image is 0.075 mJy beam$^{-1}$ and the peak surface
brightness is 25.7 mJy beam$^{-1}$. 

Figure 14: Radio images of 0744+464. At 4710 MHz the FWHM of the
restoring beam is 0.49$''$ $\times$0.44$''$, while at 8210 MHz the
beam is 0.28$''$ $\times$0.25$''$. At 4710 MHz the 
first contour level in the total
intensity image is 0.150 mJy beam$^{-1}$ and the peak surface
brightness is 120.0 mJy beam$^{-1}$. The corresponding numbers 
for the 4710 MHz polarized intensity image
are 0.12 mJy beam$^{-1}$ and 1.26 mJy beam$^{-1}$. 
At 8210 MHz the first contour level in the total
intensity image is 0.100 mJy beam$^{-1}$ and the peak surface
brightness is 58.0 mJy beam$^{-1}$. 

Figure 15: Radio images of 0748+134. At 4710 MHz the FWHM of the
restoring beam is 0.51$''$ $\times$0.45$''$, while at 8210 MHz the
beam is 0.29$''$ $\times$0.25$''$. At 4710 MHz the 
first contour level in the total
intensity image is 0.125 mJy beam$^{-1}$ and the peak surface
brightness is 3.15 mJy beam$^{-1}$. The corresponding numbers 
for the 4710 MHz polarized intensity image
are 0.12 mJy beam$^{-1}$ and 0.52 mJy beam$^{-1}$. 
At 8210 MHz the first contour level in the total
intensity image is 0.075 mJy beam$^{-1}$ and the peak surface
brightness is 0.99 mJy beam$^{-1}$. 

Figure 16: Radio images of 0828+193. At 4710 MHz the FWHM of the
restoring beam is 0.51$''$ $\times$0.46$''$, while at 8210 MHz the
beam is 0.28$''$ $\times$0.25$''$. At 4710 MHz the 
first contour level in the total
intensity image is 0.125 mJy beam$^{-1}$ and the peak surface
brightness is 6.22 mJy beam$^{-1}$. The corresponding numbers 
for the 4710 MHz polarized intensity image
are 0.12 mJy beam$^{-1}$ and 0.46 mJy beam$^{-1}$. 
At 8210 MHz the first contour level in the total
intensity image is 0.075 mJy beam$^{-1}$ and the peak surface
brightness is 2.56 mJy beam$^{-1}$. 

Figure 17: Radio images of 0943-242. At 4710 MHz the FWHM of the
restoring beam is 0.90$''$ $\times$0.44$''$, while at 8210 MHz the
beam is 0.52$''$ $\times$0.25$''$. At 4710 MHz the 
first contour level in the total
intensity image is 0.150 mJy beam$^{-1}$ and the peak surface
brightness is 40.3 mJy beam$^{-1}$. The corresponding numbers 
for the 4710 MHz polarized intensity image
are 0.12 mJy beam$^{-1}$ and 0.27 mJy beam$^{-1}$. 
At 8210 MHz the first contour level in the total
intensity image is 0.090 mJy beam$^{-1}$ and the peak surface
brightness is 13.5 mJy beam$^{-1}$. 

Figure 18: Radio images of 1106-258. At 4710 MHz the FWHM of the
restoring beam is 0.77$''$ $\times$0.44$''$, while at 8210 MHz the
beam is 0.42$''$ $\times$0.26$''$. At 4710 MHz the 
first contour level in the total
intensity image is 0.125 mJy beam$^{-1}$ and the peak surface
brightness is 28.9 mJy beam$^{-1}$. The corresponding numbers 
for the 4710 MHz polarized intensity image
are 0.12 mJy beam$^{-1}$ and 2.02 mJy beam$^{-1}$. 
At 8210 MHz the first contour level in the total
intensity image is 0.075 mJy beam$^{-1}$ and the peak surface
brightness is 11.9 mJy beam$^{-1}$. 

Figure 19: Radio images of 1113-178. At 4710 MHz the FWHM of the
restoring beam is 0.77$''$ $\times$0.44$''$, while at 8210 MHz the
beam is 0.42$''$ $\times$0.26$''$. At 4710 MHz the 
first contour level in the total
intensity image is 0.125 mJy beam$^{-1}$ and the peak surface
brightness is 20.3 mJy beam$^{-1}$. The corresponding numbers 
for the 4710 MHz polarized intensity image
are 0.12 mJy beam$^{-1}$ and 3.14 mJy beam$^{-1}$. 
At 8210 MHz the first contour level in the total
intensity image is 0.075 mJy beam$^{-1}$ and the peak surface
brightness is 6.45 mJy beam$^{-1}$. 

Figure 20: Radio images of 1138-262. At 4710 MHz the FWHM of the
restoring beam is 0.96$''$ $\times$0.44$''$, while at 8210 MHz the
beam is 0.54$''$ $\times$0.25$''$. At 4710 MHz the 
first contour level in the total
intensity image is 0.150 mJy beam$^{-1}$ and the peak surface
brightness is 32.2 mJy beam$^{-1}$. The corresponding numbers 
for the 4710 MHz polarized intensity image
are 0.12 mJy beam$^{-1}$ and 1.38 mJy beam$^{-1}$. 
At 8210 MHz the first contour level in the total
intensity image is 0.090 mJy beam$^{-1}$ and the peak surface
brightness is 7.39 mJy beam$^{-1}$. 

Figure 21: Radio images of 1232+397. At 4710 MHz the FWHM of the
restoring beam is 0.48$''$ $\times$0.44$''$, while at 8210 MHz the
beam is 0.28$''$ $\times$0.25$''$. At 4710 MHz the 
first contour level in the total
intensity image is 0.125 mJy beam$^{-1}$ and the peak surface
brightness is 12.7 mJy beam$^{-1}$. The corresponding numbers 
for the 4710 MHz polarized intensity image
are 0.12 mJy beam$^{-1}$ and 1.86 mJy beam$^{-1}$. 
At 8210 MHz the first contour level in the total
intensity image is 0.075 mJy beam$^{-1}$ and the peak surface
brightness is  4.7 mJy beam$^{-1}$. 

Figure 22: Radio images of 1324-262. At 4710 MHz the FWHM of the
restoring beam is 0.80$''$ $\times$0.44$''$, while at 8210 MHz the
beam is 0.49$''$ $\times$0.25$''$. At 4710 MHz the 
first contour level in the total
intensity image is 0.125 mJy beam$^{-1}$ and the peak surface
brightness is 54.1 mJy beam$^{-1}$. The corresponding numbers 
for the 4710 MHz polarized intensity image
are 0.12 mJy beam$^{-1}$ and 4.46 mJy beam$^{-1}$. 
At 8210 MHz the first contour level in the total
intensity image is 0.075 mJy beam$^{-1}$ and the peak surface
brightness is 25.2 mJy beam$^{-1}$. 

Figure 23: Radio images of 1345+245. At 4710 MHz the FWHM of the
restoring beam is 0.50$''$ $\times$0.49$''$, while at 8210 MHz the
beam is 0.27$''$ $\times$0.27$''$. At 4710 MHz the 
first contour level in the total
intensity image is 0.125 mJy beam$^{-1}$ and the peak surface
brightness is 71.1 mJy beam$^{-1}$. The corresponding numbers 
for the 4710 MHz polarized intensity image
are 0.12 mJy beam$^{-1}$ and 4.11 mJy beam$^{-1}$. 
At 8210 MHz the first contour level in the total
intensity image is 0.075 mJy beam$^{-1}$ and the peak surface
brightness is 29.6 mJy beam$^{-1}$. 

Figure 24: Radio images of 1410-001. At 4710 MHz the FWHM of the
restoring beam is 0.56$''$ $\times$0.45$''$, while at 8210 MHz the
beam is 0.33$''$ $\times$0.25$''$. At 4710 MHz the 
first contour level in the total
intensity image is 0.125 mJy beam$^{-1}$ and the peak surface
brightness is 16.1 mJy beam$^{-1}$. The corresponding numbers 
for the 4710 MHz polarized intensity image
are 0.12 mJy beam$^{-1}$ and 1.60 mJy beam$^{-1}$. 
At 8210 MHz the first contour level in the total
intensity image is 0.075 mJy beam$^{-1}$ and the peak surface
brightness is  6.1 mJy beam$^{-1}$. 

Figure 25: An expanded view of the hot spot region in the northwest
lobe of 1410-001. The upper image is at 4710 MHz and the lower
image is at 8210 MHz. The beams and contour levels are the same
as Figure 24.

Figure 26: Radio images of 1435+633. At 4710 MHz the FWHM of the
restoring beam is 0.51$''$ $\times$0.44$''$, while at 8210 MHz the
beam is 0.29$''$ $\times$0.25$''$. At 4710 MHz the 
first contour level in the total
intensity image is 0.125 mJy beam$^{-1}$ and the peak surface
brightness is 48.6 mJy beam$^{-1}$. The corresponding numbers 
for the 4710 MHz polarized intensity image
are 0.12 mJy beam$^{-1}$ and 0.24 mJy beam$^{-1}$. 
At 8210 MHz the first contour level in the total
intensity image is 0.075 mJy beam$^{-1}$ and the peak surface
brightness is 15.2 mJy beam$^{-1}$. 

Figure 27: Radio images of 1436+157. At 4710 MHz the FWHM of the
restoring beam is 0.49$''$ $\times$0.46$''$, while at 8210 MHz the
beam is 0.28$''$ $\times$0.26$''$. At 4710 MHz the 
first contour level in the total
intensity image is 0.180 mJy beam$^{-1}$ and the peak surface
brightness is 22.7 mJy beam$^{-1}$. The corresponding numbers 
for the 4710 MHz polarized intensity image
are 0.12 mJy beam$^{-1}$ and 2.17 mJy beam$^{-1}$. 
At 8210 MHz the first contour level in the total
intensity image is 0.100 mJy beam$^{-1}$ and the peak surface
brightness is 12.0 mJy beam$^{-1}$. 

Figure 28: Radio images of 1545-234. At 4710 MHz the FWHM of the
restoring beam is 0.76$''$ $\times$0.44$''$, while at 8210 MHz the
beam is 0.47$''$ $\times$0.26$''$. At 4710 MHz the 
first contour level in the total
intensity image is 0.125 mJy beam$^{-1}$ and the peak surface
brightness is 16.2 mJy beam$^{-1}$. The corresponding numbers 
for the 4710 MHz polarized intensity image
are 0.12 mJy beam$^{-1}$ and 0.73 mJy beam$^{-1}$. 
At 8210 MHz the first contour level in the total
intensity image is 0.075 mJy beam$^{-1}$ and the peak surface
brightness is 3.91 mJy beam$^{-1}$. 

Figure 29: Radio images of 1744+183. At 4710 MHz the FWHM of the
restoring beam is 0.49$''$ $\times$0.44$''$, while at 8210 MHz the
beam is 0.28$''$ $\times$0.25$''$. At 4710 MHz the 
first contour level in the total
intensity image is 0.125 mJy beam$^{-1}$ and the peak surface
brightness is 161  mJy beam$^{-1}$. The corresponding numbers 
for the 4710 MHz polarized intensity image
are 0.12 mJy beam$^{-1}$ and 16.0 mJy beam$^{-1}$. 
At 8210 MHz the first contour level in the total
intensity image is 0.125 mJy beam$^{-1}$ and the peak surface
brightness is 79.1 mJy beam$^{-1}$. 

Figure 30: Radio images of 1809+407. At 4710 MHz the FWHM of the
restoring beam is 0.48$''$ $\times$0.44$''$, while at 8210 MHz the
beam is 0.27$''$ $\times$0.26$''$. At 4710 MHz the 
first contour level in the total
intensity image is 0.125 mJy beam$^{-1}$ and the peak surface
brightness is 37.2 mJy beam$^{-1}$. The corresponding numbers 
for the 4710 MHz polarized intensity image
are 0.12 mJy beam$^{-1}$ and 1.09 mJy beam$^{-1}$. 
At 8210 MHz the first contour level in the total
intensity image is 0.075 mJy beam$^{-1}$ and the peak surface
brightness is 12.1 mJy beam$^{-1}$. 

Figure 31: Radio images of 1931+480. At 4710 MHz the FWHM of the
restoring beam is 0.49$''$ $\times$0.44$''$, while at 8210 MHz the
beam is 0.28$''$ $\times$0.26$''$. At 4710 MHz the 
first contour level in the total
intensity image is 0.125 mJy beam$^{-1}$ and the peak surface
brightness is 43.0 mJy beam$^{-1}$. The corresponding numbers 
for the 4710 MHz polarized intensity image
are 0.12 mJy beam$^{-1}$ and 3.03 mJy beam$^{-1}$. 
At 8210 MHz the first contour level in the total
intensity image is 0.075 mJy beam$^{-1}$ and the peak surface
brightness is 14.1 mJy beam$^{-1}$. 

Figure 32: Radio images of 2025-218. At 4710 MHz the FWHM of the
restoring beam is 0.75$''$ $\times$0.44$''$, while at 8210 MHz the
beam is 0.40$''$ $\times$0.25$''$. At 4710 MHz the 
first contour level in the total
intensity image is 0.125 mJy beam$^{-1}$ and the peak surface
brightness is 24.1 mJy beam$^{-1}$. The corresponding numbers 
for the 4710 MHz polarized intensity image
are 0.12 mJy beam$^{-1}$ and 2.97 mJy beam$^{-1}$. 
At 8210 MHz the first contour level in the total
intensity image is 0.075 mJy beam$^{-1}$ and the peak surface
brightness is 8.39 mJy beam$^{-1}$. 

Figure 33: Radio images of 2036-254. At 4710 MHz the FWHM of the
restoring beam is 0.80$''$ $\times$0.44$''$, while at 8210 MHz the
beam is 0.47$''$ $\times$0.25$''$. At 4710 MHz the 
first contour level in the total
intensity image is 0.125 mJy beam$^{-1}$ and the peak surface
brightness is 48.9 mJy beam$^{-1}$. The corresponding numbers 
for the 4710 MHz polarized intensity image
are 0.12 mJy beam$^{-1}$ and 9.35 mJy beam$^{-1}$. 
At 8210 MHz the first contour level in the total
intensity image is 0.075 mJy beam$^{-1}$ and the peak surface
brightness is 26.0 mJy beam$^{-1}$. 

Figure 34: Radio images of 2105+236. At 4710 MHz the FWHM of the
restoring beam is 0.48$''$ $\times$0.44$''$, while at 8210 MHz the
beam is 0.29$''$ $\times$0.27$''$. At 4710 MHz the 
first contour level in the total
intensity image is 0.160 mJy beam$^{-1}$ and the peak surface
brightness is 15.6 mJy beam$^{-1}$. The corresponding numbers 
for the 4710 MHz polarized intensity image
are 0.12 mJy beam$^{-1}$ and 1.64 mJy beam$^{-1}$. 
At 8210 MHz the first contour level in the total
intensity image is 0.100 mJy beam$^{-1}$ and the peak surface
brightness is 6.52 mJy beam$^{-1}$. 

Figure 35: An expanded view of the two hot spot regions in
2105+236. The upper images are at 4710 MHz and the lower
images are at 8210 MHz. The beams are the same as in Figure 34, 
and the first contour level is 0.180 mJy beam$^{-1}$ at 4710 MHz
and 0.120 mJy beam$^{-1}$ at 8210 MHz.

Figure 36: Radio images of 2139-292. At 4710 MHz the FWHM of the
restoring beam is 0.84$''$ $\times$0.44$''$, while at 8210 MHz the
beam is 0.44$''$ $\times$0.25$''$. At 4710 MHz the 
first contour level in the total
intensity image is 0.120 mJy beam$^{-1}$ and the peak surface
brightness is 43.5 mJy beam$^{-1}$. The corresponding numbers 
for the 4710 MHz polarized intensity image
are 0.12 mJy beam$^{-1}$ and 7.19 mJy beam$^{-1}$. 
At 8210 MHz the first contour level in the total
intensity image is 0.075 mJy beam$^{-1}$ and the peak surface
brightness is 16.6 mJy beam$^{-1}$. 

Figure 37: Radio images of 2141+192. At 4710 MHz the FWHM of the
restoring beam is 0.49$''$ $\times$0.44$''$, while at 8210 MHz the
beam is 0.28$''$ $\times$0.25$''$. At 4710 MHz the 
first contour level in the total
intensity image is 0.125 mJy beam$^{-1}$ and the peak surface
brightness is 34.5 mJy beam$^{-1}$. The corresponding numbers 
for the 4710 MHz polarized intensity image
are 0.12 mJy beam$^{-1}$ and 0.21 mJy beam$^{-1}$. 
At 8210 MHz the first contour level in the total
intensity image is 0.075 mJy beam$^{-1}$ and the peak surface
brightness is 14.1 mJy beam$^{-1}$. 

Figure 38: Radio images of 2202+128. At 4710 MHz the FWHM of the
restoring beam is 0.51$''$ $\times$0.44$''$, while at 8210 MHz the
beam is 0.32$''$ $\times$0.25$''$. At 4710 MHz the 
first contour level in the total
intensity image is 0.150 mJy beam$^{-1}$ and the peak surface
brightness is 19.1 mJy beam$^{-1}$. The corresponding numbers 
for the 4710 MHz polarized intensity image
are 0.12 mJy beam$^{-1}$ and 1.47 mJy beam$^{-1}$. 
At 8210 MHz the first contour level in the total
intensity image is 0.090 mJy beam$^{-1}$ and the peak surface
brightness is 7.47 mJy beam$^{-1}$. 

Figure 39: Radio images of 2251-089. At 4710 MHz the FWHM of the
restoring beam is 0.65$''$ $\times$0.44$''$, while at 8210 MHz the
beam is 0.36$''$ $\times$0.25$''$. At 4710 MHz the 
first contour level in the total
intensity image is 0.180 mJy beam$^{-1}$ and the peak surface
brightness is 45.9 mJy beam$^{-1}$. The corresponding numbers 
for the 4710 MHz polarized intensity image
are 0.12 mJy beam$^{-1}$ and 5.98 mJy beam$^{-1}$. 
At 8210 MHz the first contour level in the total
intensity image is 0.090 mJy beam$^{-1}$ and the peak surface
brightness is 20.0 mJy beam$^{-1}$. 

\vfill\eject
\bye